\documentclass[%
 aip,
 sd,%
 amsmath,amssymb,
 reprint,%
]{revtex4-1}

\usepackage{graphicx}
\usepackage{dcolumn}
\usepackage{bm}

\begin{document}

\preprint{AIP/123-QED}

\title[Breakdown in hydrogen and deuterium]{Breakdown in hydrogen and deuterium gases in static and radio-frequency fields}

\author{I. Korolov}
 \email{korolov.ihor@wigner.mta.hu}
\author{Z. Donk\'o}%
 \email{donko.zoltan@wigner.mta.hu}
\affiliation{Institute for Solid State Physics and Optics, Wigner Research Centre for Physics, Hungarian Academy of Sciences, Konkoly-Thege Miklos str. 29-33, H-1121 Budapest, Hungary
}%

\date{\today}

\begin{abstract}
We report the results of a combined experimental and modeling study of the electrical breakdown of hydrogen and deuterium in static (DC) and radio-frequency (RF, 13.56 MHz) electric fields. For the simulations of the breakdown events, simplified models are  used and only electrons are traced by Monte Carlo simulation. The experimental DC Paschen curve of hydrogen is used for the determination of the effective secondary electron emission coefficient. A very good agreement between the experimental and the calculated RF breakdown characteristics for hydrogen is found. For deuterium, on the other hand, presently available cross section sets do not allow a reproduction of RF breakdown characteristics.
\end{abstract}

\maketitle

\section{\label{sec:intro} Introduction}

Electrical breakdown in gases -- when the medium changes from an insulating phase to a conducting phase -- is a central effect in both gas discharge physics and electrical insulation.\cite{Meek1953} The physical processes and effects involved in this complex phenomenon vary to a considerable extent with the conditions, like the type and the nature of the gas, the gas pressure, the range of applied voltage (electric field) and frequency, as well as the shapes and the separation of the electrodes. The key processes in all cases are the ones directly, or indirectly responsible for the creation of charged particles: breakdown occurs when the creation of charged particles exceeds their losses.  

The minimum DC breakdown voltage between plane-parallel electrodes is typically in the domain of a few hundred Volts, and is found at pressure $\times$ electrode separation products in the $(pL)^\ast \sim $1 Torr cm range. The most famous law in the physics of gas breakdown, the Paschen law, states that the breakdown voltage ($V_{\rm B}$) depends exclusively on the parameter $pL$. This law originates from the similarity of the motion of charged particles in settings characterized by the same value of $pL$. At $pL$ values both higher and lower than $(pL)^\ast$ the breakdown voltage increases. At high $pL$ values this increase can mainly be attributed to the preference of excitation processes over ionization processes, while at low $pL$ values a higher electron energy is required to create a sufficient number of ions (and other ``active'' species) in the less frequent collisions at lower gas number densities. (We note that the principal scaling parameter is the gas number density $\times$ electrode separation product, $nL$, but it is rather convenient to use the $pL$ product instead, which is justified at a fixed temperature that links $n$ and $p$). Deviations from the Paschen law have recently  attracted considerable interest, in settings where field emission of the electrons from the electrodes play a role, e.g. in micro gaps.\cite{Klas2014,Radmilovic2013} 

While the processes involved in DC gas breakdown are quite well understood, thanks to a large number of experimental and theoretical studies for different gases and conditions\cite{Phelps1999,Hartmann2000,Klas2011,Malovic2003}, the scenarios appearing under radio frequency (RF) fields are still to be analyzed. An important aspect of RF fields is that for certain conditions charge reproduction can be achieved exclusively in the gas phase, unlike in DC discharges, where surface processes are always involved. This occurs when the electron oscillation amplitude in the RF field is smaller than the electrode gap. When the oscillation amplitude and the gap size become comparable (at low pressure and/or lower frequency), surface processes start playing a role, too, in RF fields. The changing importance of surface processes results in a particular shape of the RF breakdown curves, the characteristic features of which (e.g. turning points, in more details see later) have been addressed in several previous works.\cite{Raizer1995, Lisovkiy2008,Lisovkiy2010}

Electrical breakdown and discharges in hydrogen have been subjects of intense research due to the wide usage of this gas in a various technological processes: thin-film deposition, etching, cleaning, etc.\cite{Chu2014,Lieberman2005} The breakdown in hydrogen has been previously studied both for DC \cite{Meek1953,Carr1903, Stojanovic1990} and RF fields.\cite{Lisovskiy1998,Lisovskiy2006,Githens1940} The results obtained for RF fields are not consistent, perhaps due to the different experimental methods and conditions of these previous works. The need for solving this issue calls for well-defined measurements and accompanying simulation studies. Our aim here is to carry out such a study of the breakdown  in hydrogen, that combines experimental and kinetic simulation approaches. During the course of this work we use the same experimental setup and follow the same procedures that we have  reported in a previous study on synthetic air.\cite{Korolov2014} In our present work we also consider deuterium as working gas, to explore isotope effects on the breakdown characteristics. We note that for this gas, our literature search resulted only in DC breakdown data\cite{Armstrong} and we have not found any experimental results for RF fields. 

The paper is organized in the following way. Section II describes the experimental setup and Section III outlines the simulation techniques. In section IV we present and discuss the experimental and numerical results, while Section V gives a short summary of the work. 

\section{Experimental}

\begin{figure}[h!]
\includegraphics[width=\textwidth]{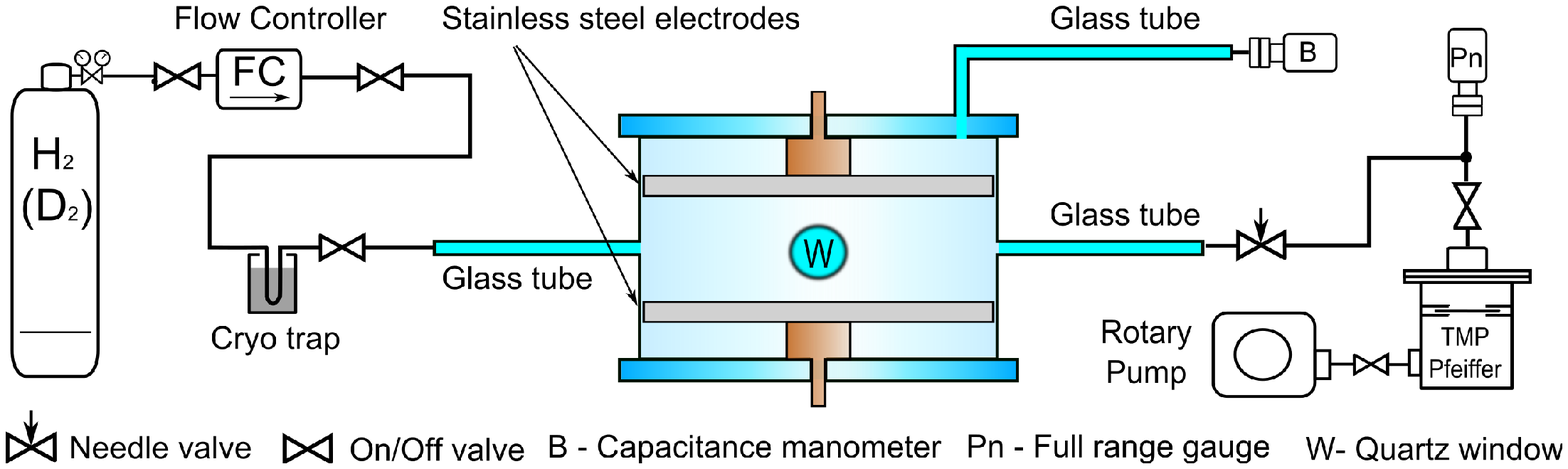}
\caption{\label{fig:experiment}Schematic of the breakdown cell and the vacuum system}
\end{figure}

The scheme of the experimental set-up is shown in figure~\ref{fig:experiment}. The geometrically symmetric discharge cell consists of two stainless steel electrodes (diameter: $D$ = 7.5 cm), which are placed inside a glass cylinder at a distance of $L$ = 1.0 cm from each other. The cell is connected to a vacuum and gas filling system using 6 mm inner diameter glass tubes (of length $\sim$35 cm). The use metallic parts near the cell is avoided in order to minimize stray capacitances that could introduce an asymmetry and, thus, influence the measurements when RF excitation is used. 

Prior to the measurements, the cell is pumped down to $<$ 10$^{-6}$ Torr by a turbomolecular pump. We use 5.0 purity hydrogen and deuterium gases (impurities according to the specification: H$_2$O and/or D$_2$O $\leq$ 3 ppm, O$_2$ $\leq$ 2 ppm, C$_n$H$_m$ $\leq$ 0.5 ppm, N$_2$ $\leq$ 5 ppm, isotopic enrichment of D$_2$ gas $\textgreater$ 99.75 atom \%) at a slow flow (2-6 sccm), regulated by a flow controller. For further purification of the gases a cryogenic trap (filled with liquid nitrogen) is used. Before starting the breakdown measurements, the electrode surfaces are cleaned by running ($\sim$8 mA) DC discharges with both polarities for about 10 minutes. To check the cleanliness of the cell, the emission spectrum  of the (cleaning) DC discharge is monitored in a wide spectral range (250-800 nm). During the measurements the temperature of the cell is kept at ambient temperature, $T \approx 300$ K.

For the DC measurements the high voltage is established by a PS325 (Stanford Research Systems) power supply, interfaced with a computer via GPIB. In the DC case two methods are applied for the determination of the breakdown voltage. In the ``voltage ramp'' (VR) method the voltage applied to the cell is slowly increased (at a rate of 0.5--1.5 V s$^{-1}$) until breakdown of the gas occurs. The rate of the voltage rise has to be low to avoid a systematic error that occurs due to the statistical nature of the breakdown event. The rate quoted above was found to be ``low enough'' to result in data that are reproducible within a few Volts. In an alternative approach the breakdown voltage is associated with the maintenance voltage of the self-sustained Townsend discharge in the limit of vanishing current.\cite{Phelps1988} For these measurements a Townsend discharge is established by connecting the DC power supply via a high resistance (106.5 M$\Omega$) to the cell. The voltage-current characteristics of the cell are recorded in the $I \sim \mu$A domain and the limit of the $V(I)$ curve at $I \rightarrow 0$ is identified as the breakdown voltage. In the following, this method is quoted as the ``low current limit'' (LCL) method. 

For the radiofrequency measurements ($f$ = 13.56 MHz) we use a ``home-made'' high-voltage push-pull oscillator\cite{Jones1997,Jones2000} based on a pair of 6146B tetrodes, which drive the electrodes with potentials $U_1 = U_0 \sin (2 \pi f t)$ and $U_2 = -U_0 \sin (2 \pi f t)$, respectively. When the breakdown event occurs, the breakdown voltage is defined as the peak value of the difference of these two potentials, $V_{\rm BR} = 2U_0$. The actual voltage on the cell is measured by two voltage divider probes (Agilent 10076B), connected to two inputs of a digitizing oscilloscope (PicoScope 6403B). The measurements are computer-controlled by LabView software. The (DC or RF) breakdown events are recognized by several methods, by detecting: (a) light emission of the plasma that forms after breakdown (using a broad band photodiode (58262 Edmund Optics)); (b) a change of the frequency of the high voltage oscillator (or sudden drop of the voltage amplitude) due a change of the impedance of the cell; (c) an increase of the power consumption of the RF oscillator. The experimental set-up and the measurement methods have been described in full details by Korolov {\it et al.} \cite{Korolov2014}

\section{Simulations}

For the major part of the conditions covered here, the motion of the charged particles, especially that of the electrons, is highly non-local\cite{nonlocal}, thus a kinetic approach has to be adopted for the description of the motion and collision processes of relevant particles. The Monte Carlo type simulation of the particles is a logical choice for this purpose. We attempt to build simplified models, which capture the most relevant processes, while neglect effects that are of secondary importance. The peculiarities of the models for the DC and RF cases are discussed below.

\subsection{DC mode}

In a DC discharge (and in DC breakdown) the major mechanisms that ensure charge reproduction are ionization and excitation processes in the gas phase and subsequent surface processes in which different species, like ions, photons, metastable atoms, and fast atoms induce the emission of secondary electrons.\cite{Phelps1999} For DC conditions surface processes play a critical role and cannot be neglected. It is noted that while the cross sections and rates of the gas phase processes are quite well known for a wide variety of gases, the roles of the different species in surface processes, at varying conditions (pressure, electric field, actual surface conditions), are largely unknown, except for certain gas-metal combinations, e.g., Ar-Cu.\cite{Phelps1999}.  

Whenever the determination of the contributions of the different ``active'' species to the secondary electron emission is unrealistic, the {\it effective secondary electron yield}, $\gamma^*$ may serve as a useful quantity; it is defined as a ratio of the electron current to ion current at the cathode and besides that of the positive ions it accounts implicitly for the contributions of the other species (fast atoms, photons, metastables) to secondary electron emission, as well. $\gamma^*$ is known to vary considerably with actual discharge and surface conditions.\cite{Hale1939, Maric2003, Donko2001, Donko2000, Malovic2003, Maric2014}

Due to the complexity of processes in H$_2$ and D$_2$ discharges we do not attempt to calculate the effective electron emission coefficient, but chose the way to determine it, as a function of $E/p$, from the {\it measured} DC Paschen curve. In our simulation code only electrons are traced using Monte-Carlo technique \cite{Tran1997,Dujko2006,Longo2011,Donko2011}, which provides a fully kinetic description under the conditions of non-local transport. We trace electrons emitted from the cathode and the additional electrons created in ionization events along their path to the anode. The cross section used for e$^-$ -- H$_2$ and e$^-$ -- D$_2$ collisions are taken from Phelps\cite{PhelpsCS} and Biagi (LxCat)\cite{CSD2}, respectively. (We note that we were not able to find any alternative, complete cross section set for D$_2$.) 

We include in the simulations the reflection of electrons from the electrode surface with a probability $\rho$. The initial number of electrons emitted from the cathode is set to $N_0 = 5\times10^3$. Some of these electrons are backscattered to the cathode and are absorbed there,\cite{Phelps1999} but for most conditions their majority creates electron avalanches. During the simulations of these avalanches we count the total number of ions created, $N_+$. We assume that all these ions reach the cathode due to their directed motion (caused by the electric field). The condition of the exact reproduction of charges that is indeed the breakdown condition, is ensured if these ions induce the emission of the same number of electrons, i.e. $N_0$, from the cathode. Consequently, the effective secondary yield is defined as 
\begin{equation}
\gamma^* = N_0 / N_+. 
\end{equation}
To obtain the effective secondary electron yield, $\gamma^*$, as a function of $E/p$, the simulations are executed for the experimentally measured values of breakdown voltage and pressure, along the Paschen curve.

\subsection{RF mode}

In contrast to the DC case, in the RF simulations the time dependent evolution of the electron density is followed. In order to greatly reduce the computational time only electrons are traced in this case, too. This simplification is justified by the fact that in the RF field the motion of ions, due to their high mass, is only slightly influenced by the time-dependent electric field, their motion can be approximated as diffusive. Considering only the lowest order diffusion mode in the cylindrical geometry of the cell, the fraction of ions, $f_i$, that reaches either of the electrodes can be estimated in a straightforward manner:\cite{Korolov2014} 
\begin{equation}
f_i =  \frac{\bigl(\frac{\pi}{L}\bigr)^2} {\bigl(\frac{\pi}{L}\bigr)^2 + \bigl(\frac{2.405}{R}\bigr)^2},
\end{equation}
where $R$ is the electrode radius. Charge reproduction at the surfaces, without tracing the ions, is modeled in the following way during the course of the simulation: whenever an ion is created at any position inside the gap, a new electron is emitted from one of the electrodes (chosen randomly) with a probability $\gamma^* f_i$ within the next RF cycle, at a random time. 

At the initialization of the simulations at a given set of conditions, we seed $5\times10^3$ electrons in the centre of the electrode gap. The simulation time is set to 500 RF cycles and the number of electrons in continuously monitored. The time-dependent electron density during the last 250 cycles is then fitted by a straight line. The slope of this line conveys information about the efficiency of charge reproduction processes: when the charge reproduction is not sufficient, the number of electrons will decrease in the simulation. An opposite behavior is found when charge production exceeds the surface losses. Separate simulations are carried out with slightly increased voltage amplitudes, and the breakdown event is recognized when the slope of the line is near zero. This algorithm closely resembles the VR (voltage-ramp) experimental approach.

\section{Results}

We discuss the experimental and simulation results for the DC and RF cases, respectively, in Sections IV.A and IV.B.

\subsection{DC mode}

Figure~\ref{fig:experiment} displays the experimental data obtained in DC mode, in comparison with data obtained in earlier works.\cite{Stojanovic1990,Meek1953,Carr1903,Githens1940} For H$_2$ we applied two approaches in the measurements of the breakdown curve: the ``voltage ramp'' (VR) method and the ``low current limit'' (LCL) method. The data obtained by these methods show a high degree of consistency. It can be clearly seen that the right (high-pressure) part of the curve is also in good agreement with results of previous works. Usage of different electrode materials and their cleanliness can explain the deviations observed in the position of the Paschen minimum, the value of the minimum breakdown voltage, and the behavior of the left (low-pressure) part of the Paschen curve. The present measurement gives a location of the Paschen minimum at $(pL)^*$ = 1.2 Torr cm with $V_{\rm BR,min}$ = 290 V, which fits well that obtained by Carr \cite{Carr1903} for brass electrodes. The largest deviation is found with the data for nickel electrodes\cite{Meek1953}, which may be attributed to different surface conditions.

The Paschen curve for D$_2$ fits well the Paschen curve of H$_2$ at high pressures. The Paschen minimum is slightly shifted towards higher $pL$, the minimum breakdown voltage is $\approx 25$V higher, and at low $pL$ we find a significantly higher DC breakdown voltage for D$_2$, compared to the corresponding values for H$_2$. The data fit reasonably well the previous experimental dataset for D$_2$.\cite{Armstrong}

\begin{figure}
\includegraphics[width=0.5\textwidth]{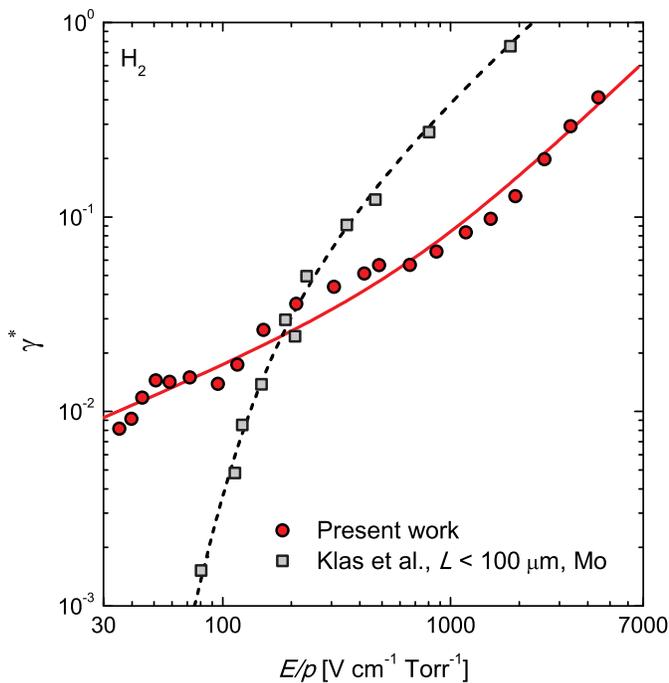}
\caption{\label{fig:gamma}Effective electron yield, $\gamma^*$, as a function of $E/p$, derived from the measured DC Paschen curve of H$_2$. For comparison we also show the data obtained for small gaps and molybdenum cathode, by Klas {\it et al.}\cite{Klas2014,Radmilovic2013}}
\end{figure}

The effective secondary electron yield for H$_2$, derived from the measured Paschen curve is shown in figure~\ref{fig:gamma} as a function of $E/p$. $\gamma^*$ is found to increase significantly towards high $E/p$ values. The present data are quite different from those obtained by Klas {\it et al.} \cite{Klas2014} for molybdenum electrodes and small, $L < 100 \mu$m, gaps. The deviation can be explained by dependence of the effective yield on the electrode material and on the gap size as at high $E/p$, e.g., field emission from the cathode cannot be neglected. (We have also computed the effective electron yield values from the D$_2$ Paschen curve, however we cannot consider these data to be reliable due to problems encountered in the simulations of the RF case, see the discussion below.)

\begin{figure}
\includegraphics[width=0.6\textwidth]{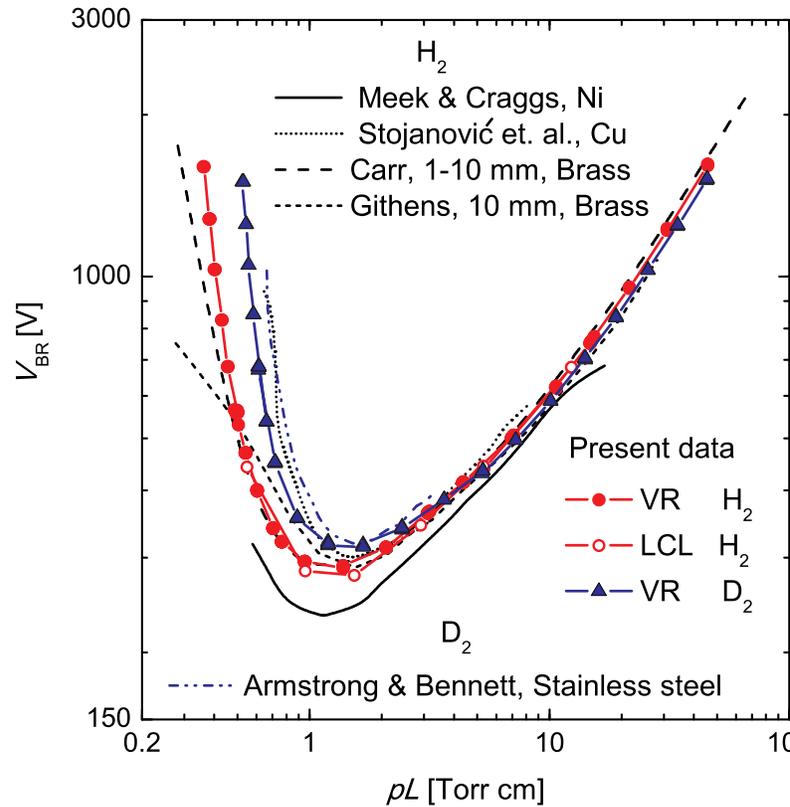}
\caption{\label{fig:DC}Breakdown voltages of H$_2$ and D$_2$ obtained in the experiments in DC mode (VR: voltage ramp and LCL: low current limit method). Paschen curves obtained in previous studies \cite{Stojanovic1990,Meek1953,Carr1903,Githens1940,Armstrong} for different electrode materials are also plotted.}
\end{figure}

\subsection{RF mode}

\begin{figure}
\includegraphics[width=0.5\textwidth]{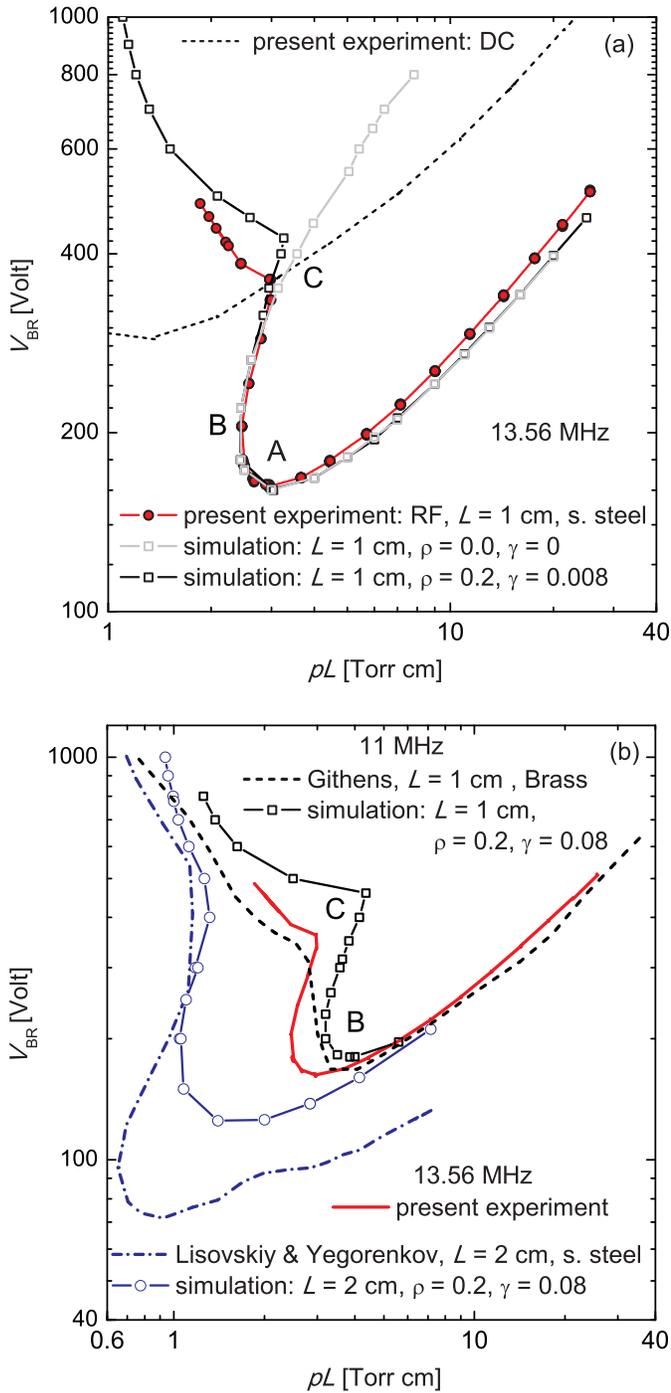}
\caption{\label{fig:RFh2sim_exp}(a) Present experimental and simulation results for RF breakdown voltage in hydrogen. ``A'', ``B'', and ``C'' are characteristic points of the breakdown curve, see text. (The DC Paschen curve, plotted with a dashed line is shown for comparison.) (b) Previous experimental results obtained by Githens\cite{Githens1940} (at $f$ = 11 MHz and $L$ = 1 cm) and Lisovskiy and Yegorenkov\cite{Lisovskiy1998} ($f$ = 13.56 MHz and $L$ = 2 cm), as well as present experimental data and simulation results for conditions of previous studies.}
\end{figure}

The measured RF breakdown curve of hydrogen gas (at $f$ = 13.56 MHz) together with our simulation results is shown in figure~\ref{fig:RFh2sim_exp}(a). The breakdown curve exhibits a particular shape. Starting from high $pL$ values, the breakdown voltage amplitude decreases towards lower $pL$ values. The minimum breakdown voltage is found at $pL \cong$ 3 Torr cm (point ``A'' in figure~\ref{fig:RFh2sim_exp}(a)). Going further along the curve, we find a turning point (point ``B'' in figure~\ref{fig:RFh2sim_exp}(a)) beyond which we experience a sudden increase of the breakdown voltage and find another turning point (point ``C'' in figure~\ref{fig:RFh2sim_exp}(a)) at $pL \cong$ 3 Torr cm and $V_{\rm BR} \cong$ 350 V, beyond which $V_{\rm BR}$ increases smoothly when $pL$ is further decreased. For the recording of most of the data points we use the voltage ramp (VR) method as described above, where the RF voltage amplitude is increased in small steps. For the conditions, where the breakdown curve becomes multivalued at fixed pressures we use the ``gas filling'' method, in which a constant RF voltage is applied to the cell and the gas pressure is slowly increased until the breakdown event occurs.\cite{Korolov2014,Lisovskiy1998}

\begin{figure}
\includegraphics[width=0.5\textwidth]{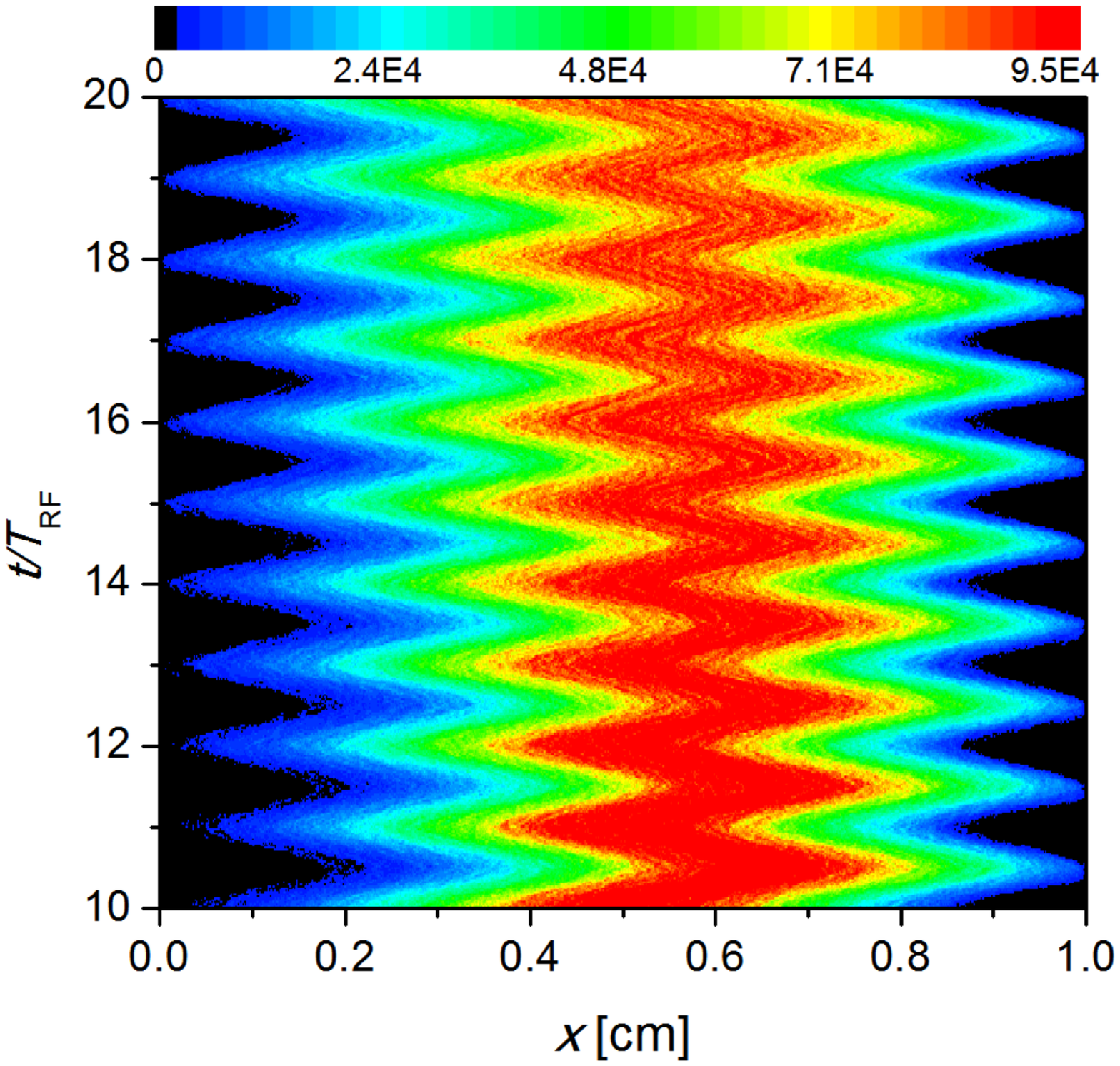}
\caption{\label{fig:xthigh} Spatio-temporal distribution of the electron density (given in arbitrary units) during 10 RF cycles (recorded from 10th RF cycle after starting the simulations) in hydrogen, at 25 Torr, 460 V. The color scale is given in arbitrary units.}
\end{figure}

\begin{figure}
\includegraphics[width=0.5\textwidth]{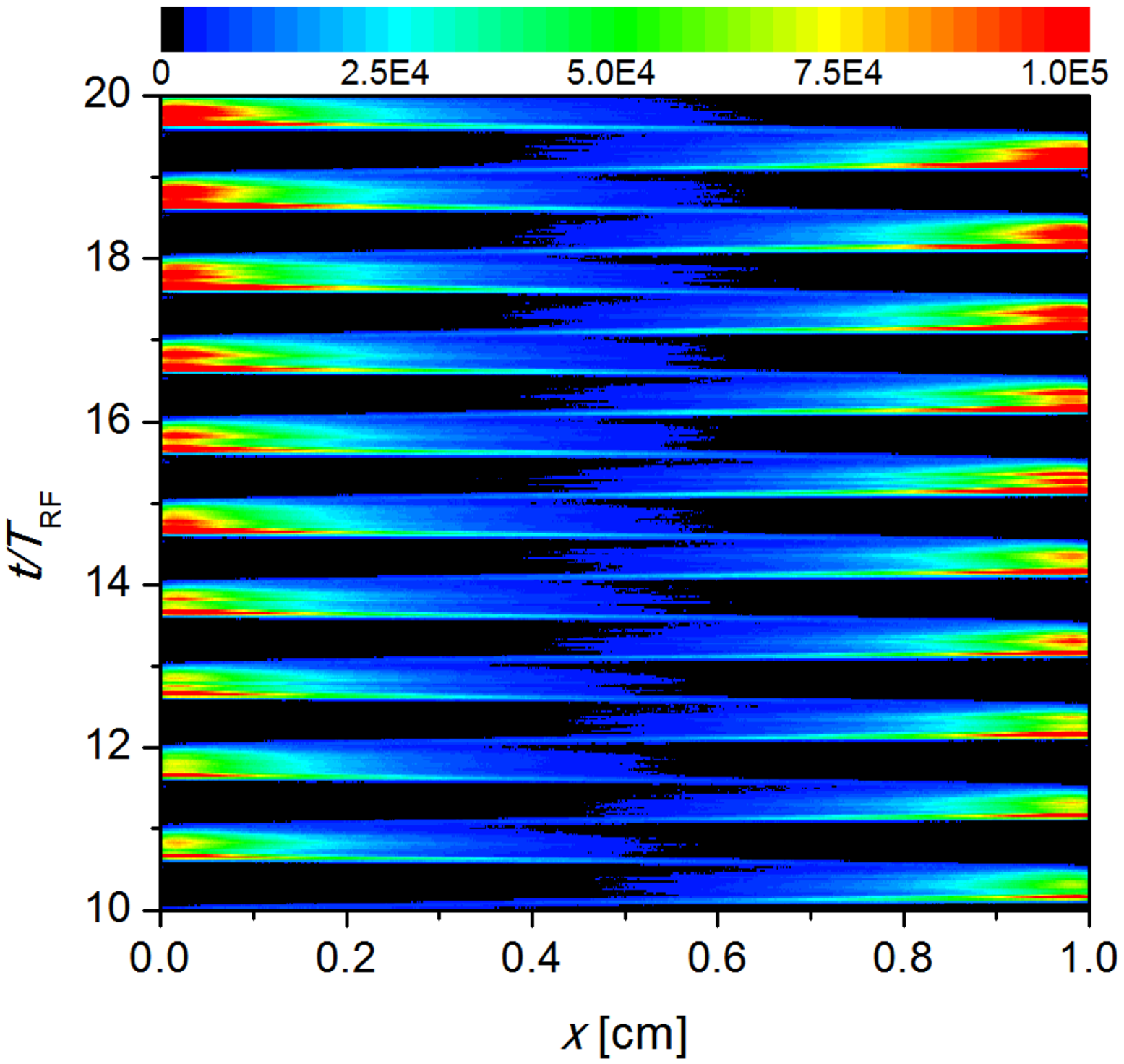}
\caption{\label{fig:xtlow}Spatio-temporal distribution of the electron density (given in arbitrary units) during 10 RF cycles (from 10th RF cycle after starting the simulations) in hydrogen at  1.2 Torr, 800 V. The color scale is given in arbitrary units.}
\end{figure}

This particular shape of the RF breakdown curve can be explained in the following way:\cite{Raizer1995, Lisovkiy2008,Lisovkiy2010} (i) at high pressures the amplitude of electron oscillations is smaller than the gap size and, therefore, the surface processes (properties of the electrodes) do not play a role. This behavior of the electrons is clearly confirmed in figure~\ref{fig:xthigh} that depicts the spatio-temporal distribution of the electron density during 10 RF cycles. (The left/right asymmetry of the plot originates from the fact that at the beginning of the simulation the electrons are seeded in the centre of the gap and are first driven to the right electrode during the first half of the RF cycle.) For a precise description of this mode only gas-phase processes have to be taken into account. This part of the curve is characterized by a breakdown voltage amplitude that decreases towards lower values of the $pL$ product.
(ii) At lower pressures, when the amplitude of the oscillations becomes comparable with the gap length, electrons can reach the surfaces where their major part gets absorbed and only their small fraction is reflected back and creates avalanches towards the opposite electrode (see fig.~\ref{fig:xtlow}).  The bending point ``C'' (see figure \ref{fig:RFh2sim_exp}(a)) at $pL \cong 3$ Torr and $V_{\rm BR}$ = 350 V corresponds to a situation where electron emission from the electrodes becomes important. 

In accordance with the above explanations, simulations that completely neglect surface effects (both secondary electron emission and electron reflection) are able to reproduce remarkably well a part of the breakdown curve, which starts at high $pL$ and ends at point ``C'' (see the grey curve in figure \ref{fig:RFh2sim_exp}(a)). This is the domain where the electron oscillation amplitude is smaller than the gap size, as explained above. However, the part of the breakdown curve beyond point ``C'' can only be reproduced by considering surface processes. As the the motion of ions in the RF field can be approximated as diffusive (as already mentioned above), the effective secondary yield $\gamma^*(E/p)$ at $E/p \rightarrow 0$ is expected to be applicable to account for surface ``creation'' of charges. The simulation results, obtained with an effective secondary electron emission yield of $\gamma^*$= 0.008 (see figure \ref{fig:gamma}) and adopting an electron reflection coefficient\cite{Kollath1956} of $\rho=0.2$ indeed reproduce the measured curve quantitatively well and confirm the existence of all the features mentioned above. This degree of agreement verifies that the model successfully captures the important physical processes, despite of the simplifications that were discussed in the previous section.

Figure~\ref{fig:RFh2sim_exp}(b) presents a comparison of the breakdown characteristics obtained in previous experimental studies of  Githens\cite{Githens1940}, as well as of  Lisovskiy and Yegorenkov\cite{Lisovskiy1998} with the present experimental data and simulation results for conditions of these previous studies. Githens\cite{Githens1940} reported data at a frequency $f$ = 11 MHz and an electrode separation of $L$ = 1 cm. These experimental data are close to our results, except that the different frequency results in a horizontal shift of the curve: at lower frequency the electron oscillation amplitude is higher at otherwise same conditions, therefore the bending point ``B'' moves to higher $pL$ at lower $f$. Our simulation results for $f$ = 11 MHz clearly reproduce the breakdown curve from high $pL$ up to this turning point. There is a deviation, however, in the location of turning point ``C'', which can be explained by the different electrode material in the experiment (brass), and by the possible onset of ion dynamics at this lower frequency (an effect that is not included in the simulation model).

The other experimental data set, by Lisovskiy and Yegorenkov\cite{Lisovskiy1998} corresponds to $f$ = 13.56 MHz and $L$ = 2 cm. The electrode material in this experiment was stainless steel, as in our work. There is a significant discrepancy between this data set and the present data, a factor of 1.8 is found (71 V vs. 125 V) between the minimum breakdown voltage values. Our simulations are not able to reproduce these experimental data.\cite{Lisovskiy1998} The deviations observed at high $pL$ values (where charge reproduction proceeds in the gas phase at small electron oscillation amplitudes) indicate some systematic error or by presence some impurities in these measurements. 

\begin{figure}
\includegraphics[width=0.5\textwidth]{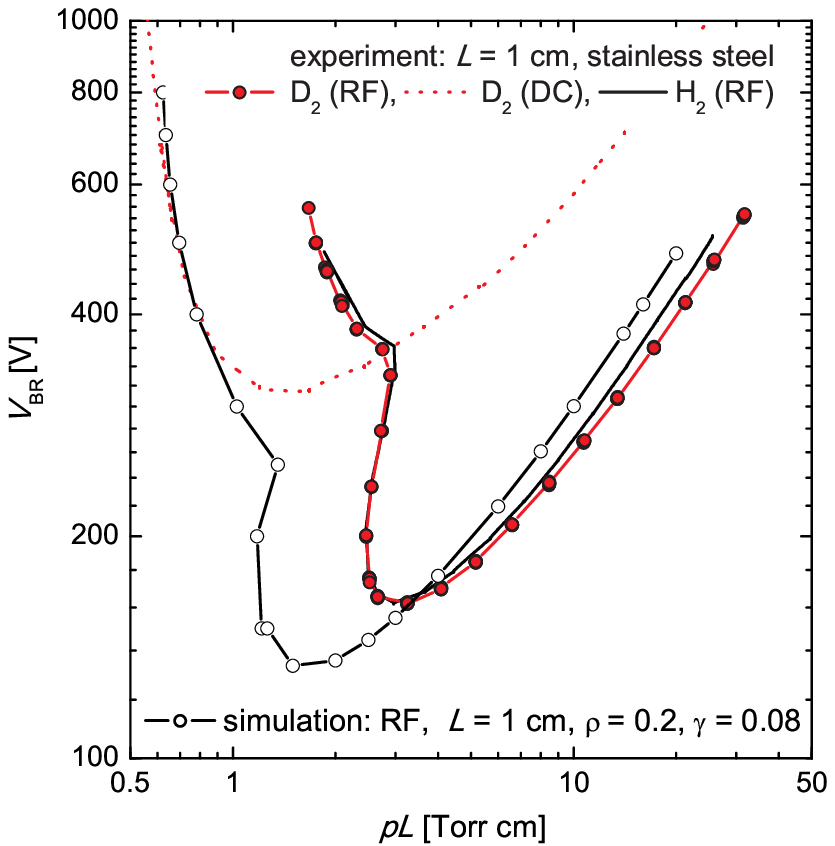}
\caption{\label{fig:RFd2sim_exp}Comparison of the experimental breakdown and simulation breakdown characteristics in D$_2$ and H$_2$.  }
\end{figure}

The breakdown characteristic for deuterium is presented in Fig.~\ref{fig:RFd2sim_exp}. The main observation is that the RF breakdown curve very nearly fits that of hydrogen. However, the simulations, with the e$^-$+D$_2$ cross section set\cite{CSD2} adopted, do not reproduce well the measured data. The deviation between these data makes this cross section set questionable and this is also the reason that we did not proceed with the determination of the effective secondary electron emission yield from the measured DC Paschen curve for D$_2$. Also, this is why the same $\gamma^\ast$ is used here in the RF simulation, as in the case of H$_2$. The simulated curve obtained for deuterium shows quite a large deviation from the measured one (see Fig.~\ref{fig:RFd2sim_exp}), at the region, where the surface processes do not play role. Therefore, the deviation can be explained by the fact that the set of the cross sections used in this simulations for deuterium is not complete, or incorrect, and has to be revised.

\section{conclusions}
We have carried out a combined experimental and kinetic simulation study of the breakdown in hydrogen and deuterium gases under DC and RF (13.56 MHz) excitation. The experiments have been performed for pressures up to 30 Torr, at $L$ = 1 cm gap length, using a glass cell, with plane parallel stainless steel electrodes. Simulation models for DC and RF conditions have been developed that trace electrons at the kinetic level. Using the model of the DC case, the effective electron emission coefficient for H$_2$ was derived from the measured Paschen curve. Despite the simplifications used in the model, the simulations showed that the whole RF breakdown curve for hydrogen can be quite accurately reproduced adopting a constant effective electron yield $\gamma^*$ = 0.008.  In case of deuterium, the deviation of the simulated breakdown curve at low pressures shows that the cross section set for e$^-$-D$_2$ collisions\cite{CSD2} should be carefully revisited.

\begin{acknowledgments}
This work has been financially supported by the Hungarian Scientific Research Fund, via Grant OTKA K 105476.
\end{acknowledgments}

\end{document}